# Observation of Brillouin scattering in a high-index doped silica chip waveguide

M. Zerbib[1], V. T. Hoang[2], J. C. Beugnot[1], K. P. Huy[1], B. Little[3], S. T. Chu[4], D. J. Moss[5], R. Morandotti[6], B. Wetzel[2], and T. Sylvestre[1,*]

1. Institut FEMTO-ST, CNRS Université de Franche-Comté, ENSMM Sup Microtech, 25030 Besançon, France
2. XLIM Research Institute, CNRS UMR 7252, Université de Limoges, 87060 Limoges, France
3. QXP Technologies Inc., Xi'an, China
4. Department of Physics, City University of Hong Kong, Tat Chee Avenue, Hong Kong, SAR, China
5. Optical Sciences Centre, Swinburne University of Technology, Hawthorn, VIC 3122, Victoria, Australia
6. INRS-EMT, 1650 Boulevard Lionel-Boulet, Varennes, J3X 1S2, Québec, Canada
*Email: thibaut.sylvestre@univ-fcomte.fr

**Abstract:** We report the observation of Brillouin backscattering in a 50-cm long spiral high-index doped silica chip waveguide and measured a Brillouin frequency shift of 16 GHz which is in very good agreement with theoretical predictions and numerical simulations based on the elastodynamics equation. © 2023

**Introduction**

Over the past decade, there has been renewed interest in the design and fabrication of integrated photonic chip waveguides to master and exploit stimulated Brillouin scattering (BS) [1]. This inelastic scattering, whereby light interacts coherently with hypersonic acoustic waves, is a powerful and flexible nonlinear optical effect for processing light and microwave, as well as an invaluable tool for the development of optical sensors, frequency combs, and lasers. While BS has been exploited earlier on within optical fibers, it is only recently that it has been demonstrated in CMOS-compatible integrated waveguides based on chalcogenide (ChG), silicon (Si), or silicon nitride (SiN) in its stoichiometric composition ($Si_3N_4$) [1-4]. Here, we demonstrate on-chip Brillouin backscattering in a 50-cm long spiral high-index (n=1.7) doped silica glass integrated waveguide [5,6].

**Experiment and Results**

The cross-section of the doped silica glass waveguide is shown in Fig. 1(a). It includes a highly-doped silica glass (HDSG) core embedded in $SiO_2$ on a SOI wafer [6]. The core has a cross-section of 1.50 μm by 1.52 μm. The chip was pigtailed at both ends using carefully aligned and UV-glued polarization-maintaining optical fibers and the total insertion loss of the 50-cm long spiral waveguide has been measured to be 8.9 dB at $\lambda$=1550 nm by using a high-resolution optical time domain reflectometer (OBR 4600 Luna Tech.). The OBR trace is shown in Fig. 1(b) for a spatial sampling resolution of 20 μm, and the linear loss has been estimated to be as low as 0.1 dB/cm. The Brillouin spectrum was then measured using a simple heterodyne technique [7], depicted in Fig. 1(c), in which the backscattered Brillouin signal from the photonic chip coherently interferes with a local oscillator and is further detected using an electrical spectrum analyzer (ESA). The resulting Brillouin spectrum is shown in Fig. 1 (d) for an input continuous-wave power of 18 dBm at 1550 nm. The Brillouin frequency shift (BFS) and its full-width at half-maximum (FWHM) linewidth were found to be $f_B$=16 GHz and $\Delta\nu_B$~350 MHz, respectively.

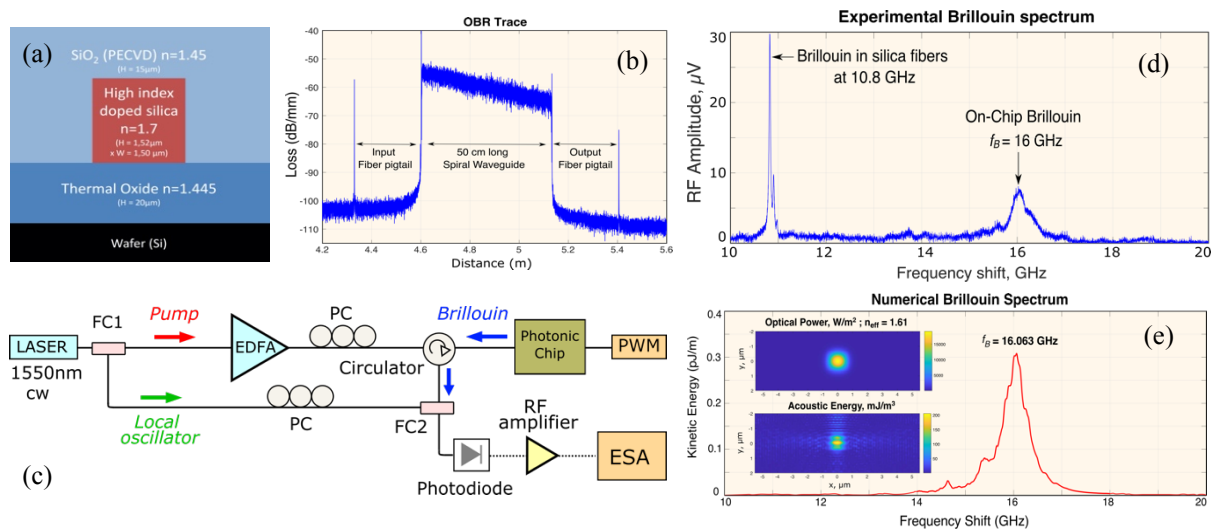

Figure 1 : (a) High-index doped silica chip waveguide cross section used for characterizing Brillouin scattering. (b) Optical backscatter reflectometer trace (spatial resolution = 20 μm). (c) Experimental setup for measuring the Brillouin backscattering spectrum. EDFA (Erbium doped fiber amplifier, PC polarization controller, FC: Fiber coupler; ESA: Electrical spectrum analyser, PWM: power meter). (d) Experimental Brillouin spectrum recorded for an input power of 18 dBm at 1550 nm, showing the Brillouin shift at 16 GHz for the high-index chip and at 11 GHz for the silica fibers in the setup. (e) Numerical simulation of the Brillouin spectrum (kinetic energy density versus acoustic frequency). The insets show the computed optical (top) and acoustic (bottom) modes.

From the Brillouin shift, we can deduce the velocity of the longitudinal acoustic wave as $V_L = f_B \lambda/(2n_{eff}) = 7700$ m.s$^{-1}$, where $n_{eff} = 1.61$ is the effective index of the fundamental optical mode, which is shown numerically in Fig. 1(e) (Top inset). This is in good agreement with theoretical prediction and numerical simulation shown in Fig. 1(d), where we plotted the computed Brillouin spectrum from the elastodynamics equation, represented as the kinetic energy density (for details about the numerical model, See Ref. [8]). The insets in Fig. 1(e) show the computed optical (top) and acoustic (bottom) transverse intensity profiles, respectively. For the optoacoustic simulation, we used the acoustic wave equation driven by the electrostrictive stress that reads as [8]

$$\rho \frac{\partial^2 u_i}{\partial t^2} + \eta_{ij} u_j - c_{ijkl} \frac{\partial^2 u_l}{\partial x_k \partial x_j} = -\chi_{klij} \epsilon_0 \frac{\partial}{\partial x_j} E_k^{(P)} E_l^{(S)*} \qquad (1)$$

where $\rho$ is the material density, $u_i$ are the displacement field components in 3D ($i \in \{x, y, z\}$), $c_{ijkl}$ are the elastic tensor components. $\eta_{ij}$ are the viscosity constants, $\chi_{kilj}$ is the electrostrictive tensor, and $\epsilon_0$ is the vacuum permittivity, respectively. $E^{(P)}$ and $E^{(S)}$ are the optical pump and Brillouin Stokes fields. For the simulation shown in Fig. 1(e), we used the parameters listed in Table 1:

Table 1

| Physical parameters | Refractive index, $n$ | Density, $\rho$ (kg/m$^3$) | Young Modulus, E (GPa) | Elastic Coefficients (GPa) | Elasto-optic Coefficients (Unitless) | Viscosity constants (Pa.s) |
|---|---|---|---|---|---|---|
| Highly-doped silica glass | 1.7 (1550 nm) | 2500 | 125.6 | $C_{11}=125.6$<br>$C_{12}=25.76$<br>$C_{44}=49.92$ | $P_{11}=0.12$<br>$P_{12}=0.27$<br>$P_{44}=-0.073$ | $\eta_{11}=2.50.10^{-2}$<br>$\eta_{12}=5.12.10^{-3}$<br>$\eta_{44}=9.93.10^{-3}$ |

**Conclusion**

In summary, we have investigated, both experimentally and theoretically, Brillouin scattering in a highly-doped silica glass photonic chip with refractive index n=1.7 and relatively low losses at 1550 nm of 0.1 dB/cm. We characterized a 50 cm long spiral waveguide, and found that the Brillouin backscattering spectrum features a high acoustic resonance near 16 GHz. We also compared the measured Brillouin frequency with theory and found very good agreement. These waveguides can be fabricated into a versatile, low-loss integrated platform, and are expected to lead to widespread Brillouin-based applications in microwave photonics [9], narrow-linewidth lasers [10], and optical frequency combs [11].

**Declaration of Competing Interest**

The authors declare that they have no known competing financial interests or personal relationships that could have appeared to influence the work reported in this paper.

**Acknowledgements**

This work has received funding from by the Agence Nationale de la Recherche (EIPHI Graduate School, ANR-17-EURE-0002, ANR-15-IDEX-03), the Bourgogne Franche-Comté Région, the European Research Council (ERC) via the European Union's Horizon 2020 research and innovation programme under grant agreement No. 950618 (STREAMLINE project). B.W. further acknowledges the support of the Conseil Régional Nouvelle-Aquitaine (SPINAL project). R.M acknowledges funding from NSERC via the Discovery and Alliance Programs, as well as from the Canada Research Chair program.